\documentclass[aps,preprintnumbers,superscriptaddress,twocolumn,amsmath,amssymb,floatfix,prl]{revtex4} 
\pdfoutput=1
\usepackage{graphicx}
\usepackage[usenames,dvipsnames]{color}
\usepackage[colorlinks,bookmarks=false,citecolor=NavyBlue,linkcolor=Red,urlcolor=blue]{hyperref}
\usepackage[export]{adjustbox}

\newcommand\be{\begin{equation}}
\newcommand\bea{\begin{eqnarray}}
\newcommand\bes{\begin{subequations}}
\newcommand\esu{\end{subequations}}
\newcommand\ee{\end{equation}}
\newcommand\eea{\end{eqnarray}}

\newcommand{\cmmnt}[1]{}

\newcommand{\dd}{\text{d}}

\def\red{\color{red}}
\def\doi{http://dx.doi.org/}

\usepackage{braket}

\newcommand\ocite[1]{[\onlinecite{#1}]}

\newcommand\titleinfo{Integrability-protected adiabatic reversibility in quantum spin chains}

\def\Tr{\operatorname{Tr}}

\begin{document}

\title{\titleinfo}

\author{Alvise Bastianello}
\affiliation{Institute for Theoretical Physics, University of Amsterdam, Science Park 904, 1098 XH Amsterdam, The Netherlands}

\author{Andrea De Luca}
\affiliation{The Rudolf Peierls Centre for Theoretical Physics, Oxford University, Oxford, OX1 3NP, United Kingdom}
%\date{\today}

\begin{abstract}
We consider the out-of-equilibrium dynamics of the Heisenberg anisotropic quantum spin--$1/2$ chain threaded by a time-dependent magnetic flux. 
In the spirit of the recently developed generalized hydrodynamics (GHD), we exploit the integrability of the model for any flux values to derive an exact description of the dynamics in the limit of slowly varying flux: the state of the system is described at any time by a time-dependent generalized Gibbs ensemble. Two dynamical regimes emerge according to the value of the anisotropy $\Delta$. For $|\Delta| > 1$, 
reversibility is preserved: the initial state is always recovered whenever the flux is brought back to zero. On the contrary, for $|\Delta| < 1$, instabilities of quasiparticles produce irreversible dynamics as confirmed by the dramatic growth of entanglement entropy. In this regime, the standard GHD description becomes incomplete and we complement it via a maximum entropy production principle. We test our predictions against numerical simulations finding excellent agreement. 
\end{abstract}


\maketitle
%\tableofcontents

Understanding the non-equilibrium dynamics of isolated many-body quantum systems is currently one of the most active research areas at the boundary between condensed matter and statistical mechanics. The importance of these studies lies in its multifaceted impact, ranging from fundamental settings, such as the microscopic derivation of thermodynamical ensembles~\cite{rigolnature, pssv, rigolreview}, to more applied ones such as the precise control of quantum systems~\cite{newExp1, grossbloch} or the realizations of novel out-of-equilibrim phases of matter~\cite{timecrystal}. In this context, cold-atom experiments have posed basic puzzles for theoretical understanding~\cite{kinoshita}, also providing a flexible playground to test and accurately validate predictions and exact results.  Quite generically, one expects that many-body systems are able to act as their own reservoirs: starting from out-of-equilibrium states $\ket{\psi}$, at long-times the expectation value of a local observable $\hat O$ approaches the thermal equilibrium one, i.e. $\bra{\psi} O(t) \ket{\psi} \to \langle O \rangle_{\mbox{\tiny eq}}$. 
This hypothesis has been thoroughly investigated in sudden quantum quenches, where an high-energy initial state $\ket{\psi}$ is evolved with a time-independent Hamiltonian $\hat H$ \ocite{calabrese-cardy}.
In practice, however, for generic systems, one has to resort to numerical simulations~\cite{vidal} which suffer by strong limitations~\cite{dmrg, altman}. For this reason, a crucial role has been played by integrable systems, for which it is possible to derive exact predictions. Several studies have clarified that integrable models which undergo a quantum quench generically exhibit relaxation~\cite{ggew4,ggew6}. However, in integrable models exist infinitely many conserved quantities $\hat{Q}_{j}=\sum_{n=1}^N\hat{q}_j(n)$ where $\hat{q}_j(n)$ is a (quasi-)local operator~\cite{lch1,lch2,lch3,lch4,lch5,lch6}.
The presence of an extensive set of integral of motions suggests that the \textit{generalised Gibbs ensemble} (GGE) $\langle \hat O \rangle_{\mbox{\tiny GGE}} = \Tr[\hat O \, e^{-\sum_j \lambda_j \hat Q_j}]/Z$ has to be used in place of the standard one~\cite{gge1, gge3}, where the appropriate set of charges has been accurately 
characterized in several studies~\cite{ggef1, ggef2, ggef3, ggef4, NESSI3}. 
The validity of the GGE conjecture has been nowadays extensively verified not only on the theoretical ground \ocite{ggew4,ggew6,ggew1,ggew2,ggew3,ggew9,ggew5,ggew7,ggew8,ggew10,lchf1,lchf2,lchf3,lchf4,lchf5}, but even on the experimental side~\ocite{ggeexp, grossbloch}.

Beyond quantum quenches, integrability constraints can be engineered to induce exotic out-of-equilibrium properties, including superdiffusive transport~\cite{NESSnum1,NESSnum2,NESSnum3, NESSf1, NESSf2, NESSf3, NESSf4, NESSf5, NESSf6, NESSf7, NESSf8, NESSf11},
dynamical ordering~\cite{quasilongrange, NESSI3} and efficient heat pumps~\cite{GGELinblad, LangeNature}. In this context, \textit{generalized hydrodynamics}~\cite{transportbertini, hydrodoyon1} (see also Ref. \ocite{GHD3,GHD6,GHD7,GHD8,GHD10,F17,DS,ID117,DDKY17,DSY17,ID217,CDV17,Kormos2018,vasseurdiff,Bas_Deluca_defhop,Bas_Deluca_defising,BDWY17,PeGa17,CDDK17,mazza2018,BFPC18,BePC18,Alba18,Bas2018,MPLC18,DeBD18,DoyonSphon17,Doyon17}) has provided a unifying framework to accurately describe integrable systems in the quasi-stationary regime which emerges from inhomogeneous initial conditions.

\begin{figure}[t!]
\includegraphics[width=0.6\columnwidth]{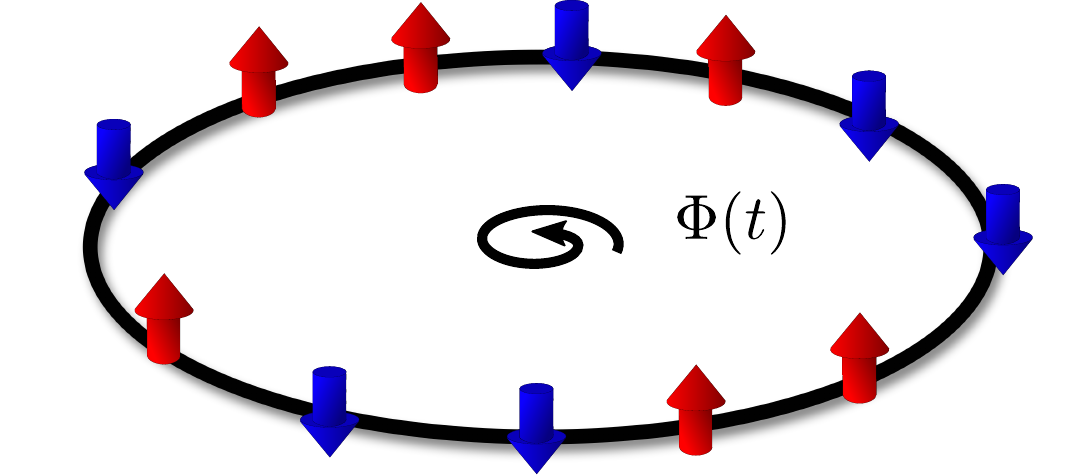}
\includegraphics[width=0.9\columnwidth]{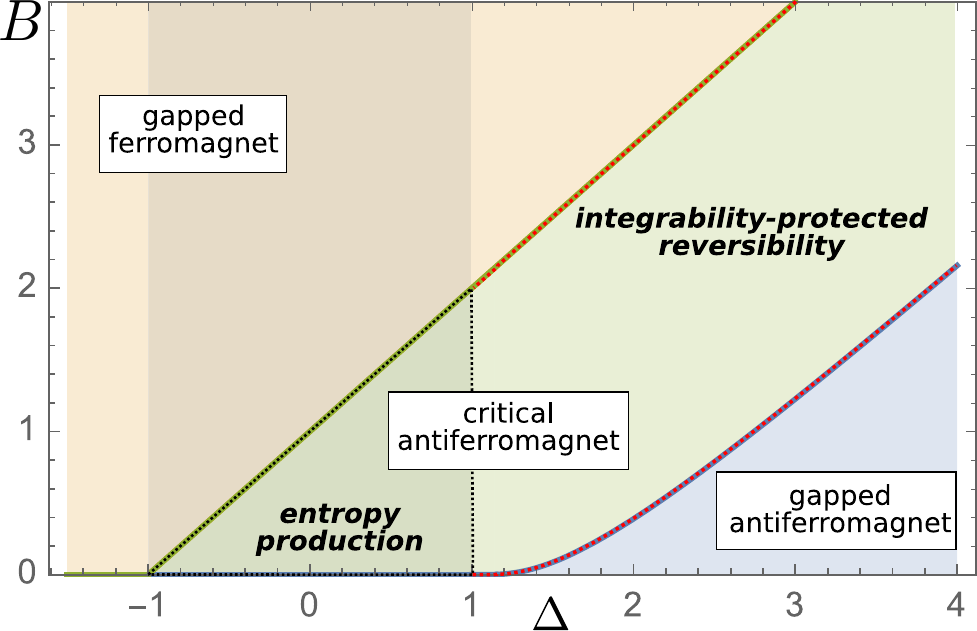}
\caption{\label{fig_phase_diag}
Top: The Heisenberg spin-$1/2$ chain is threaded by a time dependent magnetic field inducing a magnetic flux $\Phi(t)$. The instantaneous Hamiltonian~\eqref{Ham} describing the dynamics is always integrable and in the adiabatic limit, the state of the system is always locally described by a GGE ensemble.
Bottom: Phase diagram of the XXZ spin chain as a function of the anisotropy $\Delta$ and magnetic field $B$. In the region $|\Delta| \ge 1$, the model supports an infinite number of stable bound states, which preserve the reversibility of the dynamics. For $|\Delta| < 1$, the number of stable boundstates strongly depends on $\Delta$ and their momentum support does not cover the full Brillouine zone: this instability leads to irreversibility.
}
\end{figure}

In this letter we consider the out-of-equilibrium dynamics of the spin-$1/2$ XXZ chain Hamiltonian in the presence of a non-vanishing magnetic flux $\Phi$ 
\be
\label{Ham}
\hat H(\Phi) = \sum_{j=1}^N \frac{1}{2} \left(e^{\imath \Phi} \hat s_j^+ \hat s_{j+1}^- + \mbox{h.c.}\right) + \Delta \hat s_j^z \hat s_{j+1}^z  + B \hat s_j^z \, .
\ee
Above, $s^{x,y,z}$ are the usual spin$-\frac{1}{2}$ operators and $s^{\pm}=(s^x\pm i s^y)/2$, periodic boundary conditions (PBC) are enforced, together with the thermodynamic limit (TL) $N\to\infty$. By means of the Jordan-Wigner transformation, Eq.~\eqref{Ham} describes spinless fermions, where $\Delta$ controls the interaction strength and $B$ the filling. 
In this language, the flux $\Phi$ is associated with a magnetic field coupled with the $U(1)$ fermionic charge.
The system is initially prepared in an equilibrium state of the model at $\Phi=0$ (a GGE) and the flux $\Phi(t)$ is then slowly varied in time~(see Fig. \ref{fig_phase_diag} top). Infinitesimal fluxes of $|\Phi(t)| \simeq O(1/N)$ were considered in the literature in the context of linear response~\cite{ShastrySutherland1990,SutherlandAdiabatic1990} or as example of local quenches~\cite{fluxdeluca}. Sudden global quenches of the flux were considered in~\cite{prosenflux}.
Here instead, we consider finite, but slowly varying fluxes, so that the system has always time to relax to a GGE and our choice for the initial state is thus not restrictive. In particular, we focus on the groundstate (GS) of the model for different values of the anisotropy $\Delta$ and of the magnetic field $B$. 
In generic systems, slow modifications of the Hamiltonian are governed by the celebrated \emph{adiabatic theorem} \cite{adiatheo}, according to which a sufficiently slow dynamics always keeps the system in the instantaneous ground state as long as a finite gap exists. This implies reversibility of the protocol: moving slowly forth and back the external parameter, the state is back to the initial condition. However, reversibility is broken if the system is gapless: 
in the thermodynamic limit, any finite-frequency perturbation will inevitably produce excitations in the system~\cite{quantumZurek}.

However, such a picture can be drastically modified by integrability: extra dynamical symmetries
can prevent the production of excitations, even in the absence of an energy gap. While this can be expected for non-interacting systems, it leads to surprising behavior when interactions are turned on. In particular, we disclose a rich dynamical phase diagram~(see Fig. \ref{fig_phase_diag}). For $|\Delta|\ge 1$ the dynamics, within the validity of our assumptions, is fully reversible, despite the system being gapless. On the other hand, for $|\Delta|<1$ and in the gapless regime, reversibility is generally broken.
Note that the system does not stay in the instantaneous groundstate, but still, in some sense specified later on, the $\Delta-$dependent reversibility resembles the break down of the standard adiabatic theorem due to level crossing, but extended to the whole set of instantaneous conserved charges.

As the flux changes, the Hamiltonians \eqref{Ham} are connected to the $\Phi=0$ case through a gauge transformation $\hat{H}(\Phi) \sim W_\Phi^\dagger H(0)W_\Phi$, where 
boundary terms were neglected and $W_\Phi=e^{-i\Phi\sum_{j=1}^Nj \hat{s}^z_j}$. This gauge symmetry not only guarantees the instantaneous Hamiltonian to be integrable, but also connects the whole set of conserved charges for different fluxes $\hat Q_j(\Phi)=W^\dagger_\Phi \hat Q_j(0)W_\Phi$. The total magnetization $\hat S^z=\sum_{j=1}^N \hat{s}_j^z$ is flux independent and a conserved charge for any $\hat{H}(\Phi)$, thus it is constant along the time evolution.
The fact that the flux $\Phi$ can be locally (but not globally) gauged away indicates that it will not affect the microscopic scattering of quasiparticles in the model.
In the spirit of the recently introduced Generalized Hydrodynamics (GHD) \ocite{transportbertini,hydrodoyon1}, we assume a separation of time scales: the system quickly relaxes to a local GGE which slowly evolves due to the flux variation. Quantifying the precise regime of validify of hydrodynamics is an open issue~\cite{DeBD18, vasseurdiff}; here, we pragmatically assume the existence of a microscopic relaxation timescale against which the change of the flux must be compared, then benchmark our findings against numerical simulations.

\begin{figure*}[t!]
\includegraphics[width=1\textwidth]{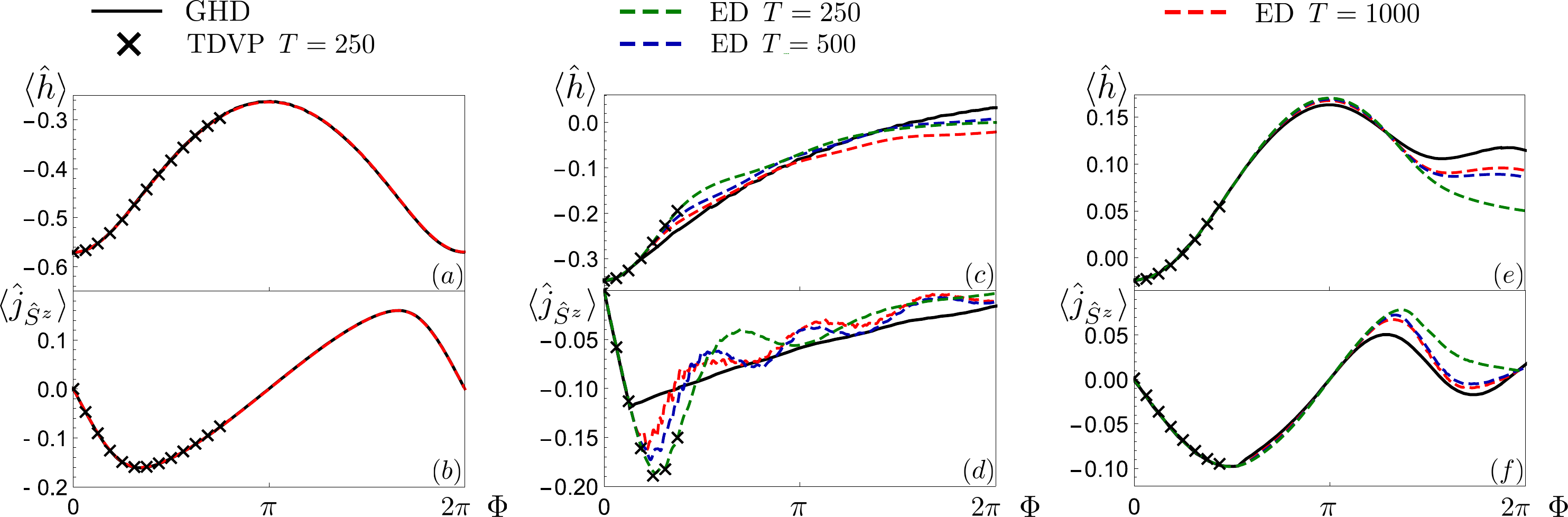}
\caption{\label{fig_2}Expectation value of the instantaneous energy and spin current vs flux. The GHD prediction is compared against ED and TDVP, the flux being changed as $\Phi(t)=2\pi t/T$. Panels a--b, we choose $\Delta=\cosh(1.5)>1$ $\langle \hat s_j^z\rangle=0.1$ (ED with $25$ sites). In this case the evolution is perfectly $2\pi-$periodic \ocite{suppl}. In b--c we rather consider $\Delta=0.5$ and $\langle \hat s_j^z\rangle=0.1$ (ED with $25$ sites), while in d--e $\langle \hat s_j^z\rangle=0.4$ (ED with $50$ sites). While for $\Delta>1$ we testified an excellent convergence even for relatively fast flux changes and small system sizes, significantly longer time scales and larger systems are needed for $\Delta<1$.
}
\end{figure*}
\paragraph{XXZ chain and generalized thermodynamics. --- }

The XXZ spin chain \eqref{Ham} is among the best known interacting integrable models. Here we provide a basic summary, leaving to the Supplementary Material (SM) \ocite{suppl} and Ref. \ocite{taka} a more exhaustive description. The spectrum of the model is dual for $\Delta\to -\Delta$, thus without loss of generality we restrict to $\Delta>0$.
Similarly to free systems, the Hilbert space of integrable models can be understood in terms of quasiparticles which undergo elastic scattering. In the presence of interactions these excitations can form bound states (also known as strings), which behave as stable quasiparticles of different species and constitute the particle content of the model.
In the thermodynamic limit, one can introduce the root densities $\rho_j (\lambda)$ which count, on average, how many quasiparticles of the species $j$ at a given rapidity $\lambda$ are present in the state. 
An exponentially large number $\sim e^{N\mathcal{S}}$ of eigenstates, \textit{microstates}, with $\mathcal{S}$ the Yang-Yang entropy~\cite{yangyang, taka}, correspond to the same macrostate identified by the root densities $\{ \rho_j(\lambda) \}_j$ and have identical local properties~\cite{quenchactionCaux}. For example, the expectation value of local charges in the thermodynamic limit (TL) is
\be\label{eq_charge}
\lim_\text{TL}N^{-1}\langle \hat Q_i\rangle=\sum_i \int \dd \lambda \, q_i^j(\lambda) \rho_j(\lambda)\, .
\ee
Above, the $\rho_j-$independent functions $q_i^j(\lambda)$ are commonly known as single-particle eigenvalues relative to the $i$-th charge and $j$-th boundstate type. Relevant examples are energy, momentum and number single-particle eigenvalue, which we indicate as $h_j(\lambda)$, $p_j(\lambda)$ and $m_j$ respectively. The number eigenvalue $m_j$ is independent of the rapidity and counts the number of fundamental particles in the boundstate and equivalently the number of spinflips.
For a complete set of charges Eq.~\eqref{eq_charge} can be inverted and in the TL, a GGE is in one-to-one correspondence with a macrostate~\cite{ggew1,lch3}.

Beyond leading to boundstates, interactions induce collective behaviors which have a net effect (\textit{dressing}) on the low-lying excitations over a GGE~\cite{bonnes, transportbertini}. 
More formally, for an arbitrary function $\tau_j(\lambda)$, we define the dressing operation as the solution of the integral linear equation
\be
\label{dresseq}
\tau^\text{dr}_j(\lambda)=\tau_j(\lambda)-\sum_{j'}\int \dd\mu \, T_{j,k}(\lambda-\mu)\sigma_{k} \vartheta_{k}(\mu)\tau_{k}^\text{dr}(\mu)\, .
\ee
where the parities $\sigma_j \in \{-1,1\}$ and the kernel $T_{j,k}(\lambda)$ depend on the value of $\Delta$.
Then, adding an excitation over GGEs modifies the charge in~\eqref{eq_charge} by a state-dependent amount governed by $(q_i^j(\lambda))^{\text{dr}}$. 

Consistently,
the density of modes available for each quasiparticle $\rho^t_j(\lambda)>\rho_j(\lambda)$ satisfies $2\pi\rho_j^t(\lambda)=\sigma_j (\partial_\lambda p_j(\lambda))^\text{dr}$.
From the filling $\vartheta_j(\lambda)=\rho_j(\lambda)/\rho_j^t(\lambda)$, one can express the Yang-Yang entropy functional $\mathcal{S}$ as
\be\label{eq_entropy}
\mathcal{S}=\sum_{j}\int \dd\lambda\, \rho_j^t \eta(\vartheta_j(\lambda))\, .
\ee
and $\eta(x)=-x\log x-(1-x)\log(1-x)$.
Eq. \eqref{eq_entropy} describes the extensive part of the entanglement entropy in a GGE state \ocite{Alba18,alba2017,alba-2018,p-18,collura-2014,bam-15,dls-13,nahum-17,nwfs-18,kauf}, 
Consistently, for groundstates $\mathcal{S}=0$, as the fillings behave as Fermi seas, i.e. $\vartheta_j(\lambda) = \delta_{j1} \Theta(\Lambda - |\lambda|)$, with $\Theta(x)$ the Heaviside function and $\Lambda$ the Fermi-point which depends on the magnetic field $B$.

\emph{The hydrodynamic approach to the flux dynamics. ---}
Let us now consider the out-of-equilibrium protocol: in particular, we imagine an infinitesimal change of the flux $\Phi\to\Phi+\dd \Phi$ and wait long enough to attain local equilibration to the new GGE. From the charge-conservation an infinite number of constraints is obtained
\be\label{eq_ch_cons}
\langle \hat Q_j(\Phi+\dd \Phi)\rangle_{\Phi+\dd \Phi}=\langle \hat Q_j(\Phi+\dd \Phi)\rangle_{\Phi}\, .
\ee

Above, with $\langle...\rangle_{\Phi}$ we mean the expectation value with respect to the GGE describing the state at flux $\Phi$. The l.h.s. of the above condition is readily computable \eqref{eq_charge}, but accessing the r.h.s. is not trivial. In this respect, the gauge transformation provides the missing information: the r.h.s. can be computed at first order in $\dd\Phi$ and, invoking the completeness of the charges, an evolution equation for $\vartheta_j(\lambda)$ can be obtained.
We leave to SM \ocite{suppl} the technical details, while here we report and comment the result: an infinitesimal increment of the flux is translated 
into a rapidity shift of the fillings, i.e.
\be\label{eq_hydro}
\vartheta_j(\lambda,\Phi+\dd \Phi) = \vartheta_j\left(\lambda-\dd\Phi\frac{m^\text{dr}_j(\lambda)}{(\partial_\lambda p_j(\lambda))^{\text{dr}}} ,\Phi\right)\, .
\ee
Let us first point out that Eq.~\eqref{eq_hydro} is formally identical to the already-known GHD equations in the presence of small force fields \ocite{GHD3}. However, its meaning is different: in Ref. \ocite{GHD3} the integrable model described by the root densities is constant in time, while in our case it is flux-dependent. We can therefore describe arbitrarily large values of $\Phi$, provided they are reached slowly enough.

The semiclassical soliton-gas interpretation~\cite{GHD6} of Eq. \eqref{eq_hydro} is clear: for any $\Phi$, $\vartheta_j(\lambda)$ describes a set of homogeneously distributed particles with momentum $p_j(\lambda)$, which undergo a collective acceleration $ p_j(\lambda) \to p_j(\lambda) + m_j \, \dd\Phi $ due to Lenz's law, i.e. the force caused by the variation of the magnetic field. Due to interactions, the effective force and momentum must be suitably dressed. 
\begin{figure}[t!]
\includegraphics[width=0.9\columnwidth]{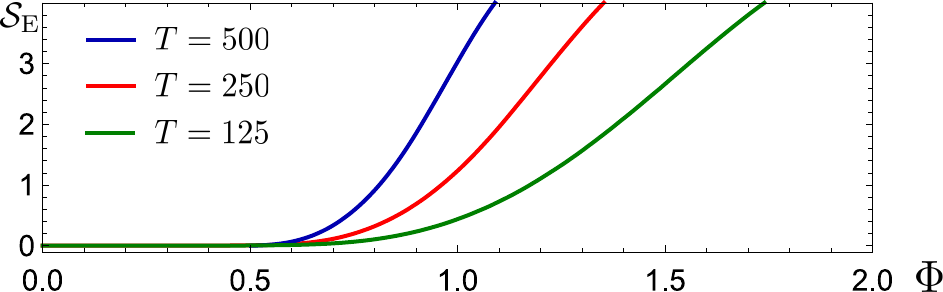}
\caption{\label{fig_entr_dmrg} Entanglement entropy relative to the initial state vs flux for $\Delta=0.5$ and $\langle \hat{s}^z_j\rangle=0.1$. The infinitely large system is bipartited in two halves and different velocities of the flux are considered $\Phi(t) = 2\pi t /T$. We interpret the plot as follows: the initial filling is a Fermi sea in the first string and a finite change in flux is needed for translating it up to the boundaries $\lambda\pm\infty$. As long as the boundaries are not involved, no entropy is produced. As soon $\lambda=\pm \infty$ is reached, the thermodynamic entropy of the GGE starts to increase, accordingly with Eq. \eqref{eq_bound_prod}. GHD predicts an infinite entanglement entropy of the infinite half as soon as $\Phi$ overcomes the critical value: consistently, the slopes in the plot increase with $T$.
}
\end{figure}
\begin{figure}[b!]
\includegraphics[width=0.9\columnwidth]{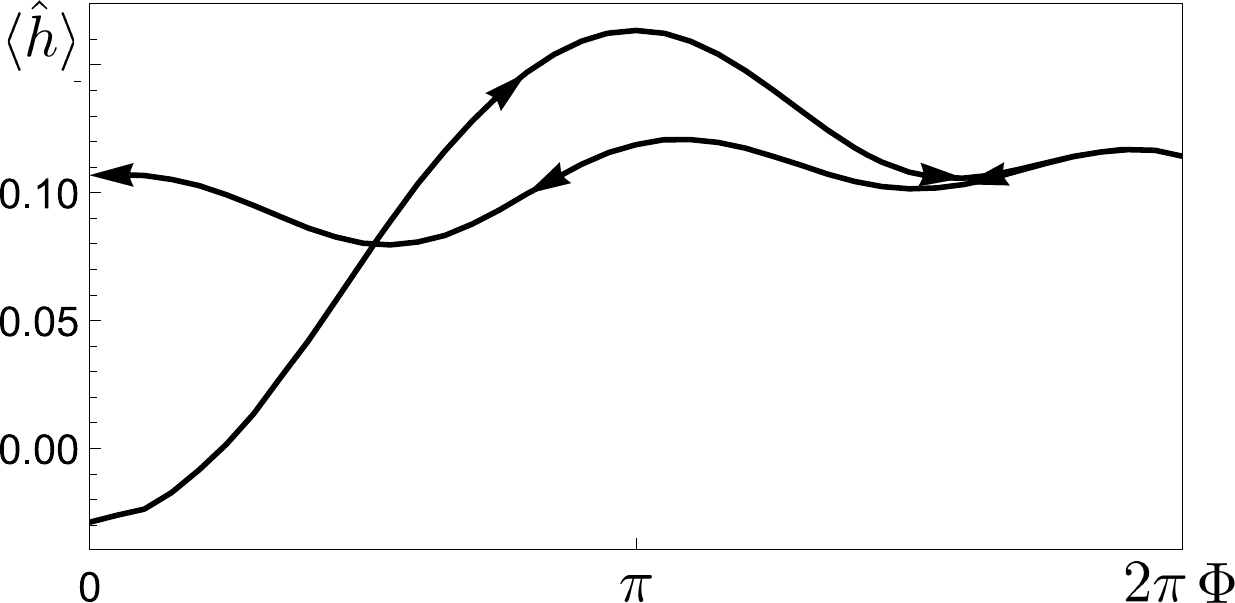}
\caption{\label{fig_irr}The irreversibility of the GHD equations is displayed for the density of energy for $\Delta=0.5$ and $\langle \hat{s}^z_j\rangle=0.4$. The flux is changed from $0$ to $2\pi$, then back.
}
\end{figure}
\paragraph{Reversible hydrodynamics for $\Delta\ge 1$.  ---}
In this case there are infinitely many strings $\{\rho_j(\lambda)\}_{j=1}^\infty$, the parity $\sigma_j=1$ in \eqref{dresseq} 
and $m_j=j$. The rapidities cover the compact domain $\lambda\in[-\pi/2,\pi/2]$ and the momenta belong to a Brillouine zone satisfying $p_j(\pi/2) = p_j(-\pi/2) \mod 2\pi$.
The kernel $T_{j,k}(\lambda)$ and single-particle eigenvalues of quasi-local charges $q_i^j(\lambda)$ are $\pi-$periodic (expressions can be found in SM \ocite{suppl}).
Periodic boundary conditions must be imposed on the fillings $\vartheta_j(\lambda)$ and thus Eqs.~\eqref{eq_hydro} guarantees the reversibility of the adiabatic protocol.
Consistently, the entropy \eqref{eq_entropy} does not change $\partial_\phi \mathcal{S}=0$ \ocite{suppl}.
In Fig. \ref{fig_2}, panels a--b, we test Eq.~\eqref{eq_hydro} against numerical simulations. We find perfect agreement.

\paragraph{Hydrodynamics for $\Delta<1$: irreversibility and entropy production.  --- }
In this case the structure of the root densities is far richer and more complicated than before \cite{taka}. The coupling is parametrized as $\Delta=\cos(\gamma)$, then the particle spectrum is finite for rational values of $\gamma/\pi$ and any value of $\Delta$ is obtained from rational approximations (see SM \ocite{suppl} for details).
In contrast with the previous case, the rapidities live on the whole real axis $\lambda\in(-\infty,\infty)$. 
In this case, $p_j(\lambda = +\infty) \neq p_j(\lambda = -\infty) \mod 2\pi$, i.e. momenta do not belong to a Brillouin zone any longer, leaving out the problem of fixing the correct boundary conditions. To clarify the issue in physical terms we resort to the semiclassic interpretation given above: increasing $\Phi$, a quasiparticle is accelerated, but for a finite increase of the flux, it reaches a momentum corresponding to infinite rapidity. What happens when the flux is further increased?
An appealing physical insight can be gained looking at single-particle eigenvalues:  since it holds true $\lim_{\lambda\to\pm\infty}q_i^j(\lambda)=m_j \times \lim_{\lambda\to\pm\infty}q_i^1(\lambda)$ \ocite{suppl}, from the point of view of any quasi-local charge, at infinite rapidity $\lambda=\pm \infty$, it is not possible to distinguish between a boundstate of type $j$ and $m_j$ unbounded excitations~\ocite{suppl}. In the absence of dynamical constraints, boundstates break and merge when $|\lambda| = \infty$: fillings shifting towards a boundary $\lambda=\pm \infty$ will recombine into fillings which emerge from the same boundary.
Being the charges unable to fix the recombination rates, we revert to the other pillar of GGE:  entropy maximization. 

Using the hydrodynamic equation, it can be showed that the change in entropy is due to boundary terms \ocite{suppl} $\partial_\Phi \mathcal{S}=\partial_\Phi \mathcal{S}^++\partial_\Phi \mathcal{S}^-$, where
\be\label{eq_bound_prod}
\partial_\Phi \mathcal{S}^{\pm}=\mp\lim_{\lambda\to\pm\infty}\left[\sum_j \sigma_j m^\text{dr}_j(\lambda) \eta(\vartheta_j(\lambda))\right]\, .
\ee
Therefore, we set as the desired boundary conditions the choice of the outgoing fillings that maximizes $\partial_\Phi \mathcal{S}^{\pm}$, together with the particle-flux conservation
$
\lim_{\lambda\to\pm\infty}\left[\sum_j    \sigma_j m_j m_j^\text{dr}(\lambda)\vartheta_j(\lambda)\right]=0\, .$
This last condition is needed to enforce Eq. \eqref{eq_ch_cons} and naturally arises in the derivation of Eq. \eqref{eq_hydro} \ocite{suppl}.
In practice, when starting from the groundstate, the Fermi sea in $\vartheta_1(\lambda)$ is shifted while the flux is increased, up to a value where the Fermi point reaches the boundary $\lambda = \infty$; then, the other fillings start to be populated according to the maximum-entropy principle. 
Our prediction is tested against numerics in Fig. \ref{fig_2}. We find good agreement although convergence to adiabaticity $\dot \Phi(t) \to 0$ is much slower than for $|\Delta|>1$. Consistently with our interpretation, the entanglement-entropy production~(see Fig. \ref{fig_entr_dmrg}) remains suppressed up to a critical value of $\Phi$ where boundstates start recombining. Then, entropy starts growing, testifying the irreversibility of the process (see Fig. \ref{fig_irr}). 
\paragraph{Conclusions. --- }
In this letter we investigated the effects of integrability on slow out-of-equilibrium protocols, focusing on the experimentally relevant case of magnetic flux in the XXZ spin chain. We unveiled the possibility of having fully reversible dynamics in a truly interacting model, despite the absence of any energy gap, as usually required by the adiabatic theorem.

However, the reversibility of the process is deeply rooted into the thermodynamic description of the system: for $|\Delta|<1$ boundstates can be recombined in an irreversible manner. We provide an hydrodynamic description of both regimes, finding good agreement with numerical simulations.
We show that GHD in the presence of force fields \ocite{GHD3} can be incomplete when lattice systems are considered due to the instability of boundstates: we complement it via maximum-entropy principle.

Finally, our numerical simulations show that, for $|\Delta| < 1$, the breaking of reversibility is associated with a very slow convergence to the hydrodynamic description. We suspect that this phenomenon is associated to a divergent relaxation timescale, similar to what happens when quantum phase transition are dynamically crossed. In this case, deviations to GHD could be universal and analogous to the Kibble-Zurek mechanism~\cite{quantumZurek}. We postpone the analysis of this intriguing possibility to future studies.

\paragraph{Acknowledgment. --- }
We are grateful to Mario Collura, Fabian Essler and Benjamin Doyon for useful discussions.
A.B. acknowledges the support from the European
Research Council under ERC Advanced grant
743032 DYNAMINT.
A.D.L. is supported by the European Union's
Horizon 2020 research and innovation programme under
the Marie Sklodowska-Curie Grant Agreement No.
794750.

%%%%%%%%%%%%%%%%%%%%%%%%%%%%%%%%%%%%%%%%%%%%%%%%%%

\onecolumngrid
%\appendix
\newpage 

\setcounter{equation}{0}            % reset equation counter
\setcounter{section}{0}             % reset section counter
\setcounter{figure}{0}             % reset figure counter
\renewcommand\thesection{\Alph{section}}    % puts letters as section numbering 
\renewcommand\thesubsection{\arabic{subsection}}    % puts letters as section numbering 
\renewcommand{\thetable}{S\arabic{table}}
\renewcommand{\theequation}{S\arabic{equation}}
\renewcommand{\thefigure}{S\arabic{figure}}
\setcounter{secnumdepth}{1}  % if the subsections need to be numbered 

\begin{center}
{\Large Supplementary Material\\ 
\titleinfo
}
\end{center}

Here we report the technical details of the results presented in the Letter. The Supplementary Material is organized as follows
\begin{enumerate}
\item Section \ref{sec_thermo}: we provide a brief summary of the Thermodynamic Bethe Ansatz description of the XXZ spin chain, further details can be found in Ref. \ocite{S_taka}.
\item Section \ref{sec_hydro_der}: we derive the hydrodynamic equation presented in the main text. For $\Delta<1$, the hydrodynamics description is best addressed through a suitable change of coordinates similarly to Ref. \ocite{S_DSY17}.
\item Section \ref{sec_num}: we present a short summary of the numerical techniques employed for a direct numerical simulation of the XXZ chain.
\end{enumerate}

\section{The XXZ spin chain and its thermodynamics}
\label{sec_thermo}

By mean of a rotation along the $x$ axis which sends $\hat{S}^x_j\to \hat{S}^x_j$, $\hat{S}^y_j\to -\hat{S}^y_j$ and $\hat{S}^z_j\to -\hat{S}^z_j$ the Hamiltonian is left unchanged, we can thus assume $\langle \hat{S}_j^z\rangle>0$ without loss of generality.
The eigenstates of the XXZ spin chain can be explicitly constructed thanks to Bethe Ansatz \cite{S_taka}. For a finite number of sites, we take as reference state that with maximum magnetization along $z$, i.e. $|0\rangle=\otimes_{i=1}^N|\uparrow_i\rangle$. Since the total spin in the $z$ direction is conserved, the eigenbasis can be organized in sector of fixed magnetization, thus we adopt a wavefunction description
\be
|\{\lambda\}_{i=1}^m\rangle=\sum_{\{j_i\}_{i=1}^m}\psi(\{j_i\}_{i=1}^m|\{\lambda\}_{i=1}^m)\hat{\sigma}^-_{j_1}...\hat{\sigma}^-_{j_m}|0\rangle\, ,
\ee
where $\hat{\sigma}_j^{-}=\hat{s}^x_j- i \hat{s}^y_j$ acts on a single site flipping down the spin: if the spin is already down, it annihilates the state.
Above, $\{\lambda\}_{i=1}^m$ are called rapidities and conveniently parametrize the state. Solving for the many-body wavefunction in generic systems is a tremendous task, but thanks to integrability a piecewise solution in terms of plane waves can be obtained
\be
\psi(\{j_i\}_{i=1}^m|\{\lambda\}_{i=1}^m)= \sum_P A(P) \prod_{j=1}^m e^{-i
	p(\lambda_{P_\ell}) j_\ell}\,,\hspace{2pc} j_1< j_2< ...< j_m\,,
\label{eq:wave_function}
\ee
Above, $p(\lambda)$ is the momentum associated with a rapidity $\lambda$. The sum is over all the permutations of the $m$ rapidities and the symmetric extension is assumed for a different ordering of the coordinates.
The coefficients for different permutations are connected through the scattering matrix: let $\Pi_{j,j+1}$ be the permutation which exchanges the rapidities at positions $j$ and $j+1$, then we have
\be
A(\Pi_{j,j+1}P)=S(\lambda_{P_j}-\lambda_{P_{j+1}})A(P)\, .
\ee
The states \eqref{eq:wave_function} are common eigenvectors of all the (quasi-)local conserved charges, which act additively on the set of the rapidities
\be
\hat Q_j|\{\lambda\}_{i=1}^m\rangle=\left(\sum_{i=1}^mq_j(\lambda_i)\right)|\{\lambda\}_{i=1}^m\rangle
\ee

Enforcing periodic boundary conditions on the wavefunction, the so called Bethe-Gaudin equations are obtained
\be
e^{-i N p(\lambda_j)}\prod_{k\ne j}S(\lambda_j-\lambda_k)=1\, ,
\ee
whose solution provides a quantization of the allowed rapidities. The model is dual for $\Delta\to -\Delta$, therefore we can focus on $\Delta>0$. However, the cases $\Delta\ge1$ and $\Delta<1$ need a separated discussion.

\subsection{Thermodynamics for $\Delta>1$}
For $\Delta\ge1$ we pose $\Delta=\cosh\theta$, then
\be
p(\lambda)=-i\log\left[\frac{\sin(\lambda-i\theta/2)}{\sin(\lambda+i\theta/2)}\right]\, \,\hspace{2pc}S(\lambda)= -\frac{\sin(\lambda+i\theta)}{\sin(\lambda-i\theta)}\, .
\ee
It should be stressed that, because of this choice in the parametrization, the rapidities naturally live within a Brillouin zone $\Re(\lambda)\in[-\pi/2,\pi/2]$.

The Bethe Gaudin equations are highly non linear and their solution is difficult. In particular, solutions with complex rapidities could exist, as it is the case for the XXZ. In the thermodynamic limit, the structure of the solution of the Bethe equations get simplified thanks to the string hypothesis \ocite{S_taka}: the solutions arrange in multiplets with the same real part of the rapidity, but shifted in the imaginary direction in a symmetric way around the real axis.
For $\Delta\ge1$ we can create multiplets of arbitrary length in the form
\be
\left\{\lambda+i\theta\frac{(M-1-2a)}{2}\right\}_{a=0}^{M-1}
\ee

These multiplets are readily interpreted as a boundstate of several particles.

Once the string hypothesis is assumed, the bound states can be considered as an unique entity labeled by the real part of the rapidities: scattering matrices for the bound states are easily extracted from the Bethe Equations. The effective momentum of a bound state is obtained summing the momenta of its components
\be
p_M(\lambda)=\sum_{a=0}^{M-1}p\left(\lambda+i\theta\frac{(M-1-2a)}{2}\right)\, ,
\ee
and similar expressions hold true for all the (quasi-)local conserved charges. In particular, we report the useful eigenvalues associated with energy and magnetization
\be
e_j(\lambda)=-\pi\sin(\theta)f_j(\lambda)\,,\hspace{2pc} m_j=j
\ee
where
\be
f_j(\lambda)=\frac{1}{\pi} \frac{\sinh (j\theta)}{\cosh(j\theta)-\cos(2\lambda)}\, .
\ee
Being interested in the thermodynamic limit, following the Thermodynamic Bethe Ansatz (TBA) recipe, we replace the fine structure of the Bethe states with counting functions: with $\rho_j(\lambda)$ we count the density of rapidities associated with the $j^\text{th}$ bound state in such a way that $L\dd\lambda \rho_j(\lambda)$ is, on average, their number in a small rapidity window around $\lambda$. 
Given a set of root densities, the expectation value of energy and magnetization easily follow
\be
\lim_{N\to\infty}N^{-1}\langle \hat{H}\rangle=\frac{\Delta}{4}+ \sum_{j=1}^\infty\int_{-\pi/2}^{\pi/2} \dd\mu \, e_j(\lambda)\rho_j(\lambda)\,,\hspace{3pc}\lim_{N\to\infty}N^{-1} \langle \hat{S}^z\rangle=\frac{1}{2}-\sum_{j=1}^\infty\int_{-\pi/2}^{\pi/2} \dd\lambda\, m_j \rho_j(\lambda)\label{s_eq_mag}\, .
\ee

Due to the necessity of microscopically satisfying the Bethe Equations, the phase space associated with each root density if modified by the presence of the other excitations
\be
\label{rhoteq}
\rho_j^t(\lambda)=f_j(\lambda)-\sum_{k=1}^\infty\int_{-\pi/2}^{\pi/2}\dd\mu\, T_{j,k}(\lambda-\mu)\rho_k(\mu)\, ,
\ee
with $2\pi f_j(\lambda)=\partial_\lambda p_j(\lambda)$ and
\be
T_{j,k}(\lambda)=(1-\delta_{j,k})f_{|j-k|}(\lambda)+f_{j+k}(\lambda)+2\sum_{\ell=1}^{\min(j,k)-{\red 1}}f_{|j-k|+2\ell}(\lambda)\, .
\ee

The phase space density is at the root of the thermodynamic of the system through the definition of the TBA entropy, which counts the number of miscroscopic states which share the same root density $\propto e^{N\mathcal{S}}$
\be
\mathcal{S}=\sum_j \int_{-\pi/2}^{\pi/2}\dd \lambda \, \rho_j^t(\lambda)\left[-\vartheta_j(\lambda)\log\vartheta_j(\lambda)-(1-\vartheta_j(\lambda))\log(1-\vartheta_j(\lambda))\right]\, ,
\ee
with the filling $\vartheta_j(\lambda)=\rho_j(\lambda)/\rho_j^t(\lambda)$.
The entropy, for a fixed rapidity window $\dd \lambda$, simply counts the (logarithm of) the number of possible arrangments of $\dd \lambda \rho_j(\lambda)$ entities in $\dd\lambda \rho_j^t(\lambda)$ empty slots.
The root densities associated with thermal states (and more general GGEs) can be derived maximizing the entropy, constrained to the knowledge of the relevant conserved charges.
For example, considering thermal states at given magnetization we are lead to the set of equations \ocite{S_taka}
\be\label{s_eq_th_TBA}
\varepsilon_j(\lambda)=\beta e_j(\lambda)+\mu m_j+\sum_{k=1}^\infty\int_{-\pi/2}^{\pi/2} \dd\mu\,  T_{j,k}(\lambda-\mu)\log(1+e^{-\beta \varepsilon_k(\mu)})\, .
\ee
Above, $\beta$ is the inverse temperature and $\mu$ the chemical potential to ensure the desired magnetization.
The function $\varepsilon_j(\lambda)$ is the effective energy parametrizing the filling through a Fermi function
\be
\vartheta_j(\lambda)=\frac{1}{1+e^{\varepsilon_j(\lambda)}}\, .
\ee
Among the possible thermal states, a relevant and simple example is the ground state at fixed magnetization: in this case, all the fillings except the first are zero, the latter being in the form of a Fermi sea
\be\label{s_eq_GS_tba}
\text{GS at fixed magnetization }\hspace{4pc}\vartheta_j(\lambda)=\begin{cases}0 &j\ne 1\\
\Theta(\Lambda-|\lambda|)\hspace{2pc}&j=1 \end{cases}
\ee
Above, $\Theta$ is the Heaviside theta function and $\Lambda$ is determined by the desired magnetization through Eq. \eqref{s_eq_mag}. 
Once the set of root densities has been computed, the expectation values of the charges easily follow, similarly to the Hamiltonian and the total magnetization \eqref{s_eq_mag}. Another quantity of interest, and particularly relevant for the forthcoming derivation of the hydrodynamic equation, is the expectation value of currents. Since all the (quasi-)local charges are conserved, their densities must satisfy proper continuity equations
\be\label{s_eq_cont}
i[\hat{q}_j(n),\hat H]+ \hat{j}_{\hat Q_j}(n+1)- \hat{j}_{\hat Q_j}(n)=0\, 
\ee
where $\hat{q}_j(n)$ is the local density of the charge $\hat Q_j$ and the current density is defined in a similar fashion
$\hat{J}_{\hat Q_j}=\sum_{n=1}^N\hat{j}_{\hat Q_j}(n)$.
Computing the expectation values of currents on arbitrary GGEs is highly non trivial and it has been an open issue until very recently \ocite{S_transportbertini,S_hydrodoyon1,S_GHD3}, when its solution laid the fundations of GHD.
In particular, for the expectation value on an arbitrary GGE it holds true
\be\label{s_eq_curr}
\langle \hat{j}_{\hat Q_j}(n)\rangle=\sum_{i=1}^\infty\int_{-\pi/2}^{\pi/2}\dd\lambda\,  \frac{(\partial_\lambda e_i(\lambda))^\text{dr}}{(\partial_\lambda p_i(\lambda))^\text{dr}}q_j^i(\lambda)\rho_i(\lambda)\, .
\ee
Above, $\frac{(\partial_\lambda e_i(\lambda))^\text{dr}}{(\partial_\lambda p_i(\lambda))^\text{dr}}$ plays the role of a (dressed) group velocity of the excitations, which then carry a charge $q_j^i(\lambda)$.
The dressing operation has been defined in the main text, but we report it hereafter for convenience for an arbitrary function $\tau_j(\lambda)$
\be
\label{dresssuppl}
\tau^\text{dr}_j(\lambda)=\tau_j(\lambda)-\sum_{j'=1}^\infty\int \dd\mu \, T_{j,j'}(\lambda-\mu)\vartheta_{j'}(\mu)\tau_{j'}^\text{dr}(\mu)\, .
\ee

\subsection{Thermodynamics for $|\Delta|<1$}

The construction of the thermodynamics for $|\Delta|<1$ closely resembles the previous case, however the string content is rather different \ocite{S_taka}. Now, we parametrize the interaction as $\Delta=\cos(\gamma)$ and set
\be\label{eq_deltaless_p}
p(\lambda)=-i\log\left[\frac{\sinh(\lambda+i\gamma/2)}{\sinh(\lambda-i\gamma/2)}\right]\, \,\hspace{2pc}S(\lambda)= -\frac{\sinh(\lambda-i\gamma)}{\sinh(\lambda+i\gamma)}\, .
\ee
In this case, the Brillouin zone for the rapidities is placed along the imaginary direction and will not appear in the thermodynamic construction. The multiplets of the string in this case have the following structure
\be
\left\{ \lambda+i\frac{\gamma}{2}(m_k+1-2a)+i\frac{\pi(1-v_k)}{4}\right\}_{a=1}^{m_k}\, ,
\ee
where the allowed magnetization eigenvalue $m_k$ and $v_k$ hugely depend on the value of $\gamma$ and will be specified later on. The integral equation defining the phase space density reads
\be
\sigma_j\rho_j^t(\lambda)=a_{m_j}^{v_j}(\lambda)-\sum_{k}\int_{-\infty}^{\infty}\dd\mu\, T_{j,k}(\lambda-\mu)\rho_k(\mu)\, .
\ee
Above, $2\pi a_{m_j}^{v_j}(\lambda)=\sigma_j\partial_\lambda p_j(\lambda)$ and the kernel is
\be
T_{j,k}(\lambda)=(1-\delta_{m_j,m_k})a_{|m_j-m_k|}^{v_j v_k}(\lambda)+a_{m_j+m_k}^{v_j v_k}(\lambda)+2\sum_{\ell=1}^{\min(m_j,m_k)-2}a_{|m_j-m_k|+2\ell}^{v_j v_k}(\lambda)\, ,
\ee
where the $a_x^y(\lambda)$ function is defined as
\be
a_{x}^{y}(\lambda)=\frac{y}{\pi}\frac{\sin(\gamma x)}{\cosh(2\lambda)-y\cos( \gamma x)}\, .
\ee
The energy eigenvalue is
\be
e_j(\lambda)=-\pi\sinh(\gamma)a_{m_j}^{v_j}(\lambda)\, .
\ee
The definition of the dressing is formally the same as in the $\Delta\ge 1$ case (apart from an extra parity $\sigma_j$, as reported in the main text) and the thermal ensembles are defined similarly to Eq. \eqref{s_eq_th_TBA}. In particular, it is still true that the GS at fixed magnetization populates only the first string in the form of a Fermi sea, similarly to Eq. \eqref{s_eq_GS_tba}. The expression for the currents is formally the same as well \eqref{s_eq_curr}.

We finally describes on the string structure, i.e. the values of $m_j, v_j$ and $\sigma_j$ for various values of the coupling. The thermodynamics is determined for rational values of the angle $\gamma/\pi$, which is parametrized in terms of continued fractions
\be
\gamma=\frac{\pi}{\nu_1+\frac{1}{\nu_2+...}}\, .
\ee
For a given continued fraction, the number of strings is $\sum_k\nu_k$. Particularly simple is the case $\gamma=\pi/\ell$, where we have
\be
m_j=j\,,\hspace{2pc} \sigma_j=1\,,\hspace{2pc} v_j=1\,\hspace{5pc} j<\ell
\ee
\be
m_\ell=1\,,\hspace{2pc} \sigma_\ell=-1\,,\hspace{2pc} v_\ell=-1
\ee

The general case is more complex. We define the auxiliary coefficients
\be
y_{-1}=0\,,\hspace{2pc} y_0=1\,, \hspace{2pc}y_1=\nu_1\,,\hspace{2pc}y_2=\nu_1\nu_2+1\,, \hspace{2pc} y_i=y_{i-2}+\nu_i y_{i-1}\, .
\ee
Then
\be
m_j=y_{i-1}+(j-s_i)y_i\, \hspace{2pc}\text{for}\,\, s_i\le j< s_{i+1}\, ,
\ee
\be
v_1=1\,, \hspace{1pc} v_{s_1}=-1\, , \hspace{2pc} v_{j\ne1}=\exp[  i\pi(m_j-1)/\gamma]\, ,
\ee
where we set $s_i=\sum_{j=1}^i \nu_j$. Finally, the parity $\sigma_j$ is $\sigma_j=(-1)^{i_j}$ where $i_j$ is defined by the inequality $s_{i_j-1}\le j<s_{i_j}$.

\section{The hydrodynamic equation}
\label{sec_hydro_der}

We now provide a careful derivation of the hydrodynamic equation presented in the main text: at the beginning, we can consider at the same time both $\Delta\ge1$ and $|\Delta|<1$.
The crucial point is computing Eq. \eqref{eq_ch_cons}, namely
\be\label{s_qe_chGGE}
\langle \hat Q_j(\Phi+\dd \Phi)\rangle_{\Phi+\dd \Phi}=\langle \hat Q_j(\Phi+\dd \Phi)\rangle_{\Phi}\, .
\ee

The l.h.s. of the above is simply computable as
\be\label{s_eq_chdphi}
\langle \hat Q_j(\Phi+\dd \Phi)\rangle_{\Phi+\dd \Phi}=N\sum_i \int \dd \lambda\, q_j^i(\lambda)\rho_i(\lambda,\Phi+\dd\Phi)\, ,
\ee
where we made explicit the flux-dependence of the root density.
In order to approach the r.h.s., we take advantage of the gauge transformation
\be
\hat Q_j(\Phi+\dd \Phi)=W^\dagger_{\dd \Phi}\hat Q_j(\Phi)W_{\dd\Phi}=\sum_n e^{-i\dd \Phi \sum_{\ell}\ell \hat{s}^z_\ell}\hat{q}_j(n,\Phi)e^{i\dd \Phi \sum_{\ell}l \hat{s}^z_\ell}
\ee

Considering $\dd \Phi$ small, we could be tempted to naively Taylor expand the above, however extra care is needed. As a matter of fact, firstly the thermodynamic limit $N\to \infty$ must be taken and only after we can consider $\dd \Phi$ to be small.
Therefore, we can proceed in the following way:
\begin{enumerate}
\item The GGE is diagonal with respect to the total magnetization $\hat{S}^z=\sum_{n=1}^N \hat{s}_n^z$, thus we have
\be
\left\langle e^{-i\dd \Phi \sum_{\ell}\ell \hat{s}^z_\ell}\hat{q}_j(n,\Phi)e^{i\dd \Phi \sum_{\ell}\ell \hat{s}^z_\ell}\right\rangle_\Phi=\left\langle e^{-i\dd \Phi \sum_{\ell}(\ell-n) \hat{s}^z_\ell}\hat{q}_j(n,\Phi)e^{i\dd \Phi \sum_{\ell}(\ell-n) \hat{s}^z_\ell}\right\rangle_\Phi
\ee

\item We now invoke the (quasi-)locality of $\hat{q}_j(n,\Phi)$ and use that its commutator with any local operator must have an exponentially vanishing support in the relative distance. In this case, and within the expectation value, we are allowed to Taylor expand the exponential for small $\dd\Phi$
\be
\left\langle e^{-i\dd \Phi \sum_{\ell}\ell \hat{s}^z_\ell}\hat{q}_j(n,\Phi)e^{i\dd \Phi \sum_{\ell}\ell \hat{s}^z_\ell}\right\rangle_\Phi=\left\langle \hat{q}_j(n,\Phi)\right\rangle_\Phi+i\dd\Phi\sum_{\ell}(\ell-n)\left\langle [\hat{s}^z_\ell,\hat{q}_j(n,\Phi)]\right\rangle_\Phi+\mathcal{O}(\dd\Phi^2)\, .
\ee
\end{enumerate}
In order to proceed further, we assume that the commutator exponentially decays with a certain characteristic length (it is the case for quasi-local charges)
\be\label{dec_com}
[\hat{s}_\ell^z,\hat{q}_j^i(n,\Phi)]\le C e^{-\xi |n-\ell|}\, .
\ee
We now regularize the expression by mean of a weight function $w_\lambda(n)=e^{-\lambda |n|}$ and consider $\mathcal{I}_\lambda$ defined as
\be
\mathcal{I}_\lambda=\frac{1}{Z}\sum_n w_\lambda(n) \sum_\ell(\ell-n)\left\langle [\hat{s}^z_\ell,\hat{q}_j(n,\Phi)]\right\rangle_\Phi\,,\hspace{3pc} Z=\sum_n w_\lambda(n)\, .
\ee
This expression is actually $\lambda-$independent, since using the translational invariance of the state we simply get $
\mathcal{I}_\lambda= \sum_l(l-n)\left\langle[\hat{s}_l^z,\hat{q}_j(n,\Phi)]\right\rangle_\Phi$.
Therefore, we now manipuate $\mathcal{I}_\lambda$ taking advantage of the limit $\lambda\to 0$ in order to get a sensible expression.
We can surely split the sum as
\be
\frac{1}{Z}\sum_n w_\lambda(n) \sum_\ell(\ell-n)\left\langle [\hat{s}^z_\ell,\hat{q}_j(n,\Phi)]\right\rangle_\Phi=\frac{1}{Z}\sum_{n,\ell} w_\lambda(n)\ell\left\langle [\hat{s}^z_\ell,\hat{q}_j(n,\Phi)]\right\rangle_\Phi-\frac{1}{Z}\sum_{n,\ell} w_\lambda(n) n\left\langle [\hat{s}^z_\ell,\hat{q}_j(n,\Phi)]\right\rangle_\Phi\, .
\ee
In the first term, thanks to the exponential decay of the commutator \eqref{dec_com}, we can replace $w_\lambda(n)\to w_\lambda(\ell)$ introducing an error $\mathcal{O}(\lambda)$
\be
\frac{1}{Z}\sum_{\ell} w_\lambda(l) l\left\langle \left[\hat{s}^z_\ell,\sum_n\hat{q}_j(n,\Phi)\right]\right\rangle_\Phi-\frac{1}{Z}\sum_{n} w_\lambda(n) n\left\langle \left[\sum_\ell \hat{s}^z_\ell,\hat{q}_j(n,\Phi)\right]\right\rangle_\Phi+\mathcal{O}(\lambda)\, .
\ee

We now use that the commutators are nothing else than the action of an extensive charge over a local charge density. We already wrote the continuity equation coming from the action of the Hamiltonian on a charge density, but in integrable models all the conserved charges are on the same footing. Eq. \eqref{s_eq_cont} can be indeed generalized to
\be
i[\hat{q}_i(n),\hat Q_{i'}]+ \hat{j}_{\hat Q_i}^{\hat Q_{i'}}(n+1)- \hat{j}_{\hat Q_i}^{\hat Q_{i'}}(n)=0\, ,
\ee
where
$\hat{J}_{\hat Q_i}^{\hat Q_{i'}}=\sum_{n=1}^N\hat{j}_{\hat Q_i}^{\hat Q_{i'}}(n)$.
Thus
\be
\mathcal{I}_\lambda=\frac{i}{Z}\sum_{\ell} w_\lambda(\ell) \ell\left\langle\Bigg(\hat{j}_{\hat S^z}^{ \hat Q_{j}(\Phi)}(\ell+1)-\hat{j}_{\hat S^z}^{ \hat Q_{j}(\Phi)}(\ell)\Bigg)\right\rangle_\Phi+\frac{i}{Z}\sum_{n} w_\lambda(n) n\left\langle\Bigg( \hat{j}^{\hat S^z}_{ \hat Q_{j}(\Phi)}(n+1)-\hat{j}^{\hat S^z}_{ \hat{Q}_{j}(\Phi)}(n)\Bigg)\right\rangle_\Phi+\mathcal{O}(\lambda)\, .
\ee
We can use once more translational invariance and organize the above as
\be
\mathcal{I}_\lambda=\frac{i}{Z}\sum_{n}\big[ w_\lambda(n-1) (n-1)-w_\lambda(n)n]\Bigg(\langle\hat{j}_{\hat S^z}^{ \hat{Q}_{j}(\Phi)}(n)\rangle_\Phi+\langle \hat{j}^{\hat S^z}_{ \hat{Q}_{j}(\Phi)}(n)\rangle_\Phi\Bigg)+\mathcal{O}(\lambda)\, .
\ee
Above, boundary terms are safely neglected thanks to the exponential decay of $w_\lambda(n)$. Now, at the price of another inessential $\mathcal{O}(\lambda)$ correction we can replace $w_\lambda(n-1)\to w_\lambda(n)$ and finally find

\be
\mathcal{I}_\lambda=-i\Bigg(\langle\hat{j}_{\hat S^z}^{ \hat{Q}_{j}(\Phi)}(n)\rangle_\Phi+\langle \hat{j}^{\hat S^z}_{ \hat{Q}_{j}(\Phi)}(n)\rangle_\Phi\Bigg)+\mathcal{O}(\lambda)\, .
\ee
As we already noticed, the value of $\mathcal{I}_\lambda$ is actually $\lambda-$independent, thus we can take $\lambda\to 0$ and simply drop the $\mathcal{O}(\lambda)$ term, finally reaching the equality
\be
\langle \hat Q_j(\Phi+\dd \Phi)\rangle_\Phi=\langle \hat Q_j(\Phi)\rangle_\Phi+\dd\Phi\Bigg(\langle\hat{J}_{\hat S^z}^{ \hat{Q}_{j}(\Phi)}\rangle_\Phi+\langle \hat{J}^{\hat S^z}_{ \hat{Q}_{j}(\Phi)}\rangle_\Phi\Bigg)\, .
\ee

The expectation of the generalized currents on arbitrary GGEs can be computed similarly to Eq. \eqref{s_eq_curr} \ocite{S_GHD3}
\be
\lim_{N\to\infty}N^{-1}\langle \hat J_{ \hat Q_{i}(\Phi)}^{\hat Q_j(\Phi)}\rangle_\Phi=\sum_l \int \dd \lambda \, q^l_j(\lambda) \frac{(\partial_\lambda q_i^l(\lambda))^\text{dr}}{(\partial_\lambda p_l(\lambda))^\text{dr}}\rho_l(\lambda,\Phi)\, .
\ee
Notice that $\langle\hat{J}_{\hat S^z}^{ \hat{Q}_{j}(\Phi)}\rangle_\Phi=0$.
Combining now these last results together with \eqref{s_qe_chGGE} \eqref{s_eq_chdphi} we get an infinite set of integral equations that the root densities must satisfy

\be
\sum_i \int \dd \lambda\, q^i_j(\lambda)\partial_\Phi\rho_i(\lambda,\Phi)=\sum_i \int \dd \lambda \, m_i \frac{(\partial_\lambda q_j^i(\lambda))^\text{dr}}{(\partial_\lambda p_i(\lambda))^\text{dr}}\rho_i(\lambda,\Phi)\, .
\ee

We can further simplify the above noticing that $(\partial_\lambda p_i(\lambda))^\text{dr}=2\pi \sigma_i \rho_i^t(\lambda)$ (in the $\Delta\ge 1$ case, we conventionally set $\sigma_j=1$), then the above can be rewritten as
\be\label{s_eq_hydroint}
\sum_i \int \dd \lambda\, q^i_j(\lambda)\partial_\Phi\rho_i(\lambda,\Phi)=\sum_i \int \frac{\dd \lambda}{2\pi} \,\sigma_i m_i (\partial_\lambda q_j^i(\lambda))^\text{dr}\vartheta_i(\lambda,\Phi)\, .
\ee

Taking advantage of the symmetry of the kernel $T_{j,j'}$, it holds true the following identity for arbitrary functions $\tau_i(\lambda)$ and $\alpha_i(\lambda)$
\be
\sum_i \int \dd \lambda \,\sigma_i \tau_i(\lambda) \alpha_i^\text{dr}(\lambda)\vartheta_i(\lambda)=\sum_i \int \dd \lambda \,\sigma_i \tau^\text{dr}_i(\lambda) \alpha_i(\lambda)\vartheta_i(\lambda)\, .
\ee

Applying this to \eqref{s_eq_hydroint} we readily get
\be\label{s_eq_hydroint2}
\sum_i \int \dd \lambda\, q^i_j(\lambda)\partial_\Phi\rho_i(\lambda,\Phi)=\sum_i \int \frac{\dd \lambda}{2\pi}\partial_\lambda q_j^i(\lambda) \,\sigma_i m^\text{dr}_i(\lambda) \vartheta_i(\lambda,\Phi)\, .
\ee

Our final goal is to extract the hydrodynamic equation from the above, invoking the completeness of the set of charges. Hereafter, it is convenient to consider separately the case $\Delta\ge1$ and $|\Delta|<1$.

\subsection{The hydrodynamics for $\Delta\ge1$}

We start from \eqref{s_eq_hydroint2} and perform an integration by parts. Since $\lambda$ lives in a Brillouin zone, the boundary terms vanish and we simply get
\be
\sum_i \int_{-\pi/2}^{\pi/2} \dd \lambda\, q^i_j(\lambda)\partial_\Phi\rho_i(\lambda,\Phi)=-\sum_i \int_{-\pi/2}^{\pi/2} \frac{\dd \lambda}{2\pi} q_j^i(\lambda) \partial_\lambda\Big( m^\text{dr}_i(\lambda) \vartheta_i(\lambda,\Phi)\Big)\, .
\ee

Now, in the same spirit of the original derivation of GHD \ocite{S_transportbertini,S_hydrodoyon1}, we invoke the completeness of the set of the conserved charges: from the infinite set of integral equations we conclude the validity of a differential equation for the root density
\be
\label{hydrorho}
\partial_\Phi\rho_i(\lambda,\Phi)=-\frac{1}{2\pi}  \partial_\lambda\Big( m^\text{dr}_i(\lambda) \vartheta_i(\lambda,\Phi)\Big)\, .
\ee
As the last passage, manipulating the integral equations defining $\rho_i^t$ similarly to the derivation presented in \ocite{S_transportbertini,S_GHD3}, the final desired expression can be reached
\be\label{s_eq_hydro_fill}
\partial_\Phi\vartheta_i(\lambda,\Phi)=-\frac{m^\text{dr}_i(\lambda)}{2\pi\rho_i^t(\lambda)} \partial_\lambda \vartheta_i(\lambda,\Phi)\, ,
\ee

which also implies
\be
\partial_\Phi\rho_i^t(\lambda,\Phi)=-\frac{1}{2\pi}\partial_\lambda m_i^\text{dr}(\lambda)\, .
\ee

It is not difficult to prove that these hydrodynamic equations preserve the Yang-Yang entropy ($\eta(x)=-x\log x-(1-x)\log(1-x)$)
\be
\mathcal{S}=\sum_i\int_{-\pi/2}^{\pi/2} \dd\lambda\, \rho_i^t(\lambda,\Phi) \eta(\vartheta_i(\lambda,\Phi))\, ,
\ee
following the derivation of Ref. \ocite{S_CDDK17}.
Indeed, taking its flux derivative we have
\be
\partial_\Phi \mathcal{S}=\sum_i\int_{-\pi/2}^{\pi/2} \dd\lambda\, \partial_\Phi\rho_i^t(\lambda,\Phi) \eta(\vartheta_i(\lambda,\Phi))+\sum_i\int_{-\pi/2}^{\pi/2} \dd\lambda\, \rho_i^t(\lambda,\Phi) \partial_\Phi \vartheta_i(\Lambda,\Phi) \eta'(\vartheta_i(\lambda,\Phi))\,,
\ee
\begin{figure}[t!]
\includegraphics[width=0.7\columnwidth]{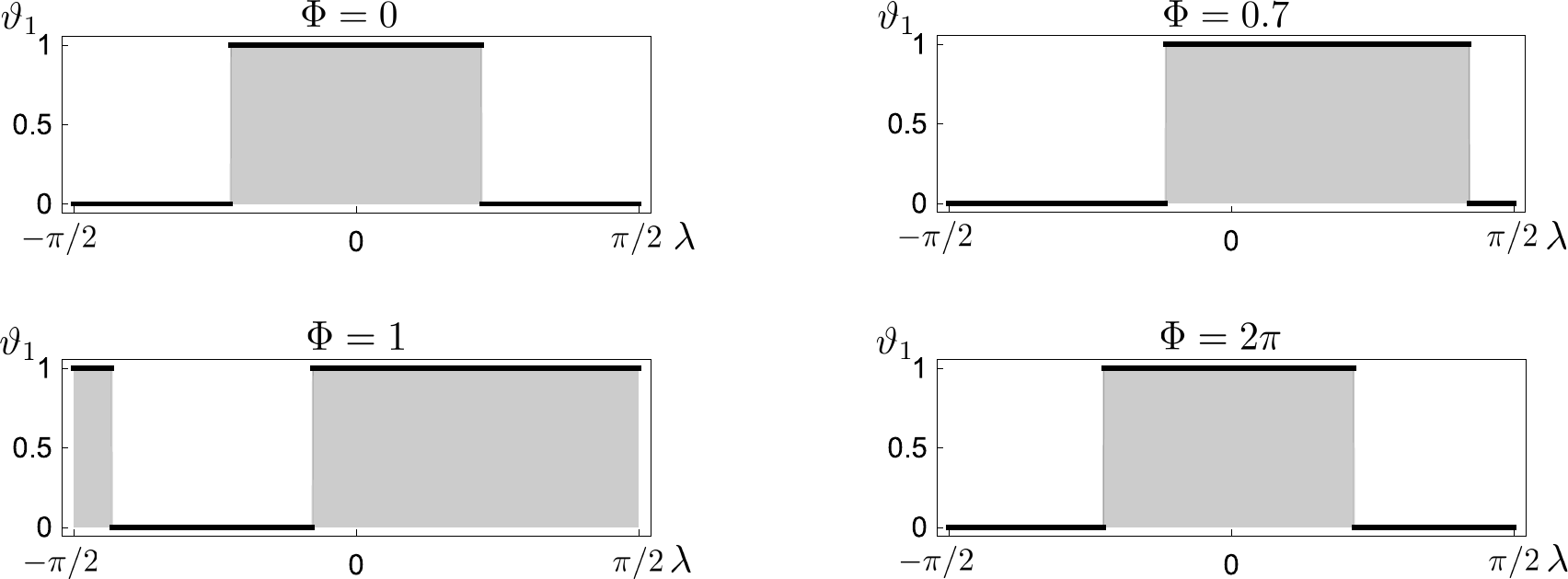} 
\caption{\label{fig_filling_deltalarge}Example of the flux evolution of the filling for $\Delta\ge1$. The system is initialized in the GS with magnetization $\langle \hat{s}_j^z\rangle=0.1$ and the interaction is chosen $\Delta=\cosh(1.5)$. Only the first string is populated and, as the flux changes, the Fermi sea rigidly shifts. The evolution is $2\pi$ periodic (see also Section \ref{sec_periodic}).
}
\end{figure}
then, using the hydrodynamic equations we rewrite the above as
\begin{multline}
\partial_\Phi \mathcal{S}=-\frac{1}{2\pi}\sum_i\int_{-\pi/2}^{\pi/2} \dd\lambda\, \partial_\lambda m^\text{dr}_i(\lambda) \eta(\vartheta_i(\lambda,\Phi))-\frac{1}{2\pi}\sum_i\int_{-\pi/2}^{\pi/2} \dd\lambda\,m^\text{dr}_i\partial_\lambda \vartheta_i \eta'(\vartheta_i(\lambda,\Phi))=\\
-\frac{1}{2\pi}\int_{-\pi/2}^{\pi/2}\dd \lambda\, \partial_\lambda \Big(m^\text{dr}_i(\lambda)\eta(\vartheta_i(\lambda,\Phi))\Big)\, .
\end{multline}
Using the periodicity at the boundaries of the Brillouin zone, the last integral is zero.
The hydrodynamic equations are most easily numerically solved in terms of infinitesimal shifts, namely we rewrite \eqref{s_eq_hydro_fill} as
\be
\vartheta_i(\lambda,\Phi+\dd \Phi)= \vartheta_i\left(\lambda-\dd\Phi\frac{m^\text{dr}_i(\lambda)}{2\pi\rho_i^t(\lambda)} ,\Phi\right)\, 
\ee

and periodic boundary conditions are imposed at the boundaries $\pm \pi/2$.
The rapidity space is discretized as well and the dressing equations are experienced to converge through a simple iteration scheme. As we already discussed, for $\Delta\ge1$ an infinite number of root densities exists, therefore generic GGEs must necessarily be approximated by mean of a truncation of the number of strings. However, the ground state populates only the first string in the form of a Fermi sea: since for $\Delta\ge1$ no mixing among the strings is present, if the system is initialized in the ground state, then the hydrodynamic solution remains a shifted Fermi sea in the first string, zero in the others. We plot an explicit example in Fig. \ref{fig_filling_deltalarge}.

\subsection{The hydrodynamics for $|\Delta|<1$}

In this case we proceed from Eq. \eqref{s_eq_hydroint2} by mean of an integration by parts as well, however since we do not have any Brillouin zone, the resulting boundary terms cannot be neglected
\be\label{s_eq_hydro_bound}
\sum_i \int_{-\infty}^\infty \dd \lambda\, q^i_j(\lambda)\partial_\Phi\rho_i(\lambda,\Phi)=-\sum_i \int_{-\infty}^\infty \frac{\dd \lambda}{2\pi} q_j^i(\lambda) \,\partial_\lambda\Big(\sigma_i m^\text{dr}_i(\lambda) \vartheta_i(\lambda,\Phi)\Big)+\left[\sum_i \frac{1}{2\pi} q_j^i(\lambda) \sigma_i m^\text{dr}_i(\lambda) \vartheta_i(\lambda,\Phi)\right]_{\lambda=-\infty}^{\lambda=+\infty}\, .
\ee
Invoking the completeness of the charges, we can still get the  ``bulk" hydrodynamic equations
\be
\partial_\Phi\rho_j(\lambda,\Phi)=- \frac{1}{2\pi}\,\partial_\lambda\Big(\sigma_j m^\text{dr}_j(\lambda) \vartheta_j(\lambda,\Phi)\Big)\, .
\ee
This equation, as for the $\Delta\ge 1$, can be rewritten in the desired form in terms of the filling
\be\label{q_eq_hydrodeltaless}
\partial_\Phi \vartheta_j(\lambda)=-\frac{\sigma_j m_j^\text{dr}(\lambda)}{2\pi \rho_j^t(\lambda)}\partial_\lambda \vartheta_j(\lambda)\, .
\ee
For $|\Delta|<1$ there is not a Brillouin zone, therefore the choice of the correct boundary conditions at $\lambda=\pm\infty$ is less evident. From the boundary terms of Eq. \eqref{s_eq_hydro_bound} further constraints follow, but we should first understand the behavior of the (quasi-)local charges as $\lambda\to\pm\infty$.

As we stated in Section \ref{sec_thermo}, the eigenvalue associated with a given string is obtained summing over the components of the bound state, namely
\be\label{eq_q_sum}
q_\ell^{j}(\lambda)=\sum_{a=1}^{m_j}q_\ell^1\left( \lambda+i\frac{\gamma}{2}(m_j+1-2a)+i\frac{\pi(1-v_j)}{4}\right)\, .
\ee

The rapidities are only a convenient parametrization: the true variables characterizing a state are the momenta, as it is evident from the Bethe wavefunction \eqref{eq:wave_function}. Therefore, the charge eigenvalue for the $1^\text{st}$ string (which is the fundamental excitation) is not actually a function of the rapidity, but is rather a smooth function of the momentum $p(\lambda)$ \eqref{eq_deltaless_p} (and of course periodic in the real part, which can be thought to live in the Brillouin zone $\Re(p(\lambda))\in(-\pi,\pi)$). Thus $q_\ell^{1}(\lambda)=\tilde{q}_\ell^1(p(\lambda))$ with $\tilde{q}_\ell^1$ smooth. Therefore, we rewrite Eq. \eqref{eq_q_sum} as
\be\label{eq_q_sum}
q_\ell^{j}(\lambda)=\sum_{a=1}^{m_j}\tilde{q}_\ell^1\left(p\left( \lambda+i\frac{\gamma}{2}(m_j+1-2a)+i\frac{\pi(1-v_j)}{4}\right)\right)\, .
\ee

Now, we are ready to take the $\lambda\to\pm \infty$ limit. It is immediate to notice that, regardless the imaginary part of the rapidity, we have
\be
\lim_{\lambda\to\pm\infty}p\left( \lambda+i\frac{\gamma}{2}(m_j+1-2a)+i\frac{\pi(1-v_j)}{4}\right)=\pm  \gamma
\ee
and this immediately tells us
\be\label{eq_q_sum}
\lim_{\lambda\to\pm\infty}q_\ell^j(\lambda)=m_j\tilde{q}_\ell^1\left(\pm\gamma\right)\, .
\ee

Therefore, at the boundaries, the charge eigenvalues of the various strings are proportional to the same quantity: there are no charges that can distinguish among a bound state of $m_j$ excitations or a set of $m_j$ unbounded excitations.

This implies that from the boundary terms of \eqref{s_eq_hydro_bound} we can extract at most one global conservation law for each boundary
\be\label{s_eq_magflux}
\sum_i  m_i \sigma_i m^\text{dr}_i(\lambda) \vartheta_i(\lambda,\Phi)\Big|_{\lambda=+\infty}=0\,,\hspace{4pc}\sum_i  m_i \sigma_i m^\text{dr}_i(\lambda) \vartheta_i(\lambda,\Phi)\Big|_{\lambda=-\infty}=0
\ee

Of course, a single constraint is in general not enough to fix the proper boundary conditions.

As we already commented in the main text, we revert to the thermodynamic entropy to fix the remaining boundary conditions. Indeed, the foundations of the GGE are based on fixing the expectation value of all the relevant charges, together with a maximization of entropy: if there are not charges able to distinguish among different root densities at $\lambda=\pm\infty$, then only with the maximum entropy recipe is left.
As for the $\Delta\ge1$ case, we compute the variation of the entropy $\partial_\Phi \mathcal{S}$, reaching
\be
\partial_\Phi \mathcal{S}=
-\frac{1}{2\pi}\int_{-\infty}^{\infty}\dd \lambda\, \partial_\lambda \Big( \sigma_im^\text{dr}_i(\lambda) \eta(\vartheta_i(\lambda,\Phi))\Big)\, .
\ee

This time the boundary contributions cannot be discarded and thus we get $\partial_\Phi\mathcal{S}=\partial_\Phi \mathcal{S}^++\partial_\Phi\mathcal{S}^-$ where
\be\label{s_eq_entropy_rate}
\mathcal{S}^\pm=\mp \frac{1}{2\pi}\Big( \sigma_i m^\text{dr}_i(\lambda) \eta(\vartheta_i(\lambda,\Phi))\Big)_{\lambda=\pm\infty}\, .
\ee

Therefore, in order to identify the proper boundary conditions, we impose maximum entropy production together with the constraints \eqref{s_eq_magflux}.
In order to discuss how this recipe can be implemented in practice, it is useful to rewrite the hydrodynamics equations in terms of an implicit change of variables, as we now discuss.

\subsubsection{Implementing the hydrodynamic equations: a change of variables}

The hydrodynamic equations for $|\Delta|<1$ are not easily implemented in the rapidity space: indeed, for large rapidities, $\rho_i^t$ vanishes and thus the force term in Eq. \eqref{q_eq_hydrodeltaless} diverges. As a matter of fact, thanks to the divergence of the force, the filling at a given $\lambda$ is shifted to infinite rapidities for a finite change of the flux.
Closely following Ref. \ocite{S_GHD3}, we then introduce a change of variables $\lambda\to \tau$ where, for any different root density, we define
\be\label{s_eq_tauvar}
\tau_j(\lambda)=2\pi \int_{-\infty}^\lambda\dd\mu\, \rho_j^t(\mu,\Phi)\, .
\ee
The change of variables is of course flux dependent.
Being $\rho_j^t$ strictly positive, the relation can be inverted and the function $\tau_j$ lives within the interval $[0,L_j(\Phi)]$, where we define
\be
L_j(\Phi)=2\pi\int_{-\infty}^\infty\dd\mu\, \rho_j^t(\mu,\Phi)\, .
\ee
The length of the intervals $L_j$ changes with the value of the flux. Indeed, we immediately get
\be
\partial_\Phi L_j(\Phi)=\sigma_j (m^\text{dr}(-\infty)-m^\text{dr}(+\infty))\, .
\ee
We then define the fillings in the new coordinates $\tilde{\vartheta}_j$ by mean of the equality
\be
\tilde{\vartheta}_j(\tau_j(\lambda),\Phi)=\vartheta_j(\lambda,\Phi)\, .
\ee
In terms of the new variables, the hydrodynamic equations assume a remarkably simple form
\be\label{s_eq_hydrotau}
\partial_\Phi \tilde{\vartheta}_j(\tau,\Phi)=-\sigma_j m^\text{dr}_j(-\infty) \partial_\tau \tilde{\vartheta}_j(\tau,\Phi)\, .
\ee

Thus, the fillings in the $\tau$ space rigidly translate, the direction being decided by the sign of $\sigma_j m^\text{dr}_j(-\infty)$.
In principle, we should go forth and back from the $\tau$ space to the rapidity one in order to solve for the dressed magnetization, however due to the fact that $\partial_\lambda \tau_j(\lambda)$ vanishes at $\lambda\to\pm\infty$, the fillings are smoothed more and more in the rapidity space as they flow towards infinity. Thus, in the dressing operation we can look at the fillings as if they were constant and this greatly simplifies the computation of the dressed magnetization, which is reduced to a set of algebraic equations

\begin{figure}[t!]
\includegraphics[width=1\columnwidth]{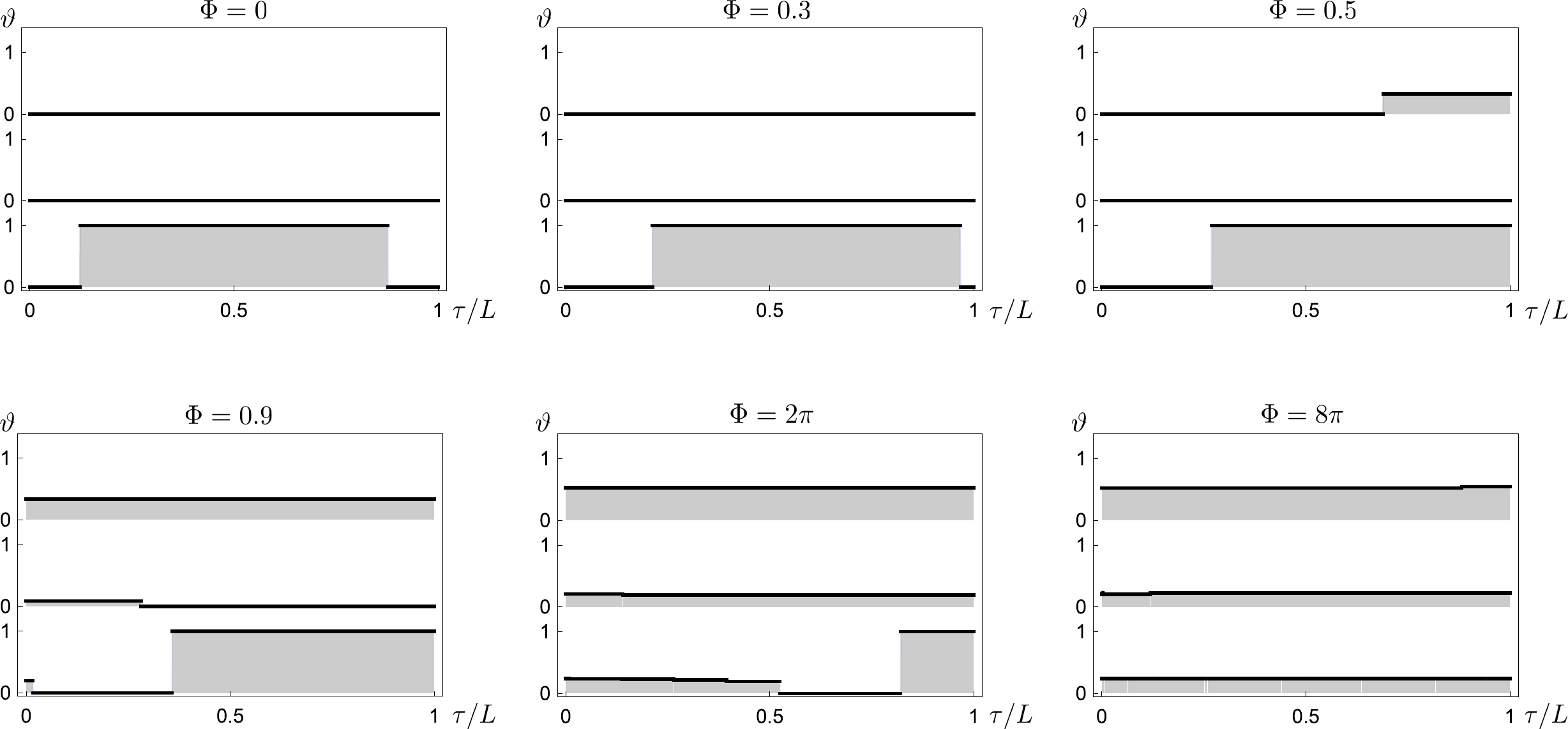} 
\caption{\label{fig_filling_deltaless}Example of the flux evolution of the fillings for $\Delta=1/2$, starting from the GS with magnetization $\langle \hat{s}_j^z\rangle=0.1$, the flux is steadily increased. For $\Delta=1/2$, three root densities are present, in each plot we depict all the strings (ordered from the bottom to the top). On the horizontal axis we rescaled the coordinate $\tau/L_j$ for each string, in such a way the definition domain is always $[0,1]$. When a Fermi sea hits the boundary, it splits in the other strings following the maximization is the entropy creation rate. For large values of the flux, the root densities appear to be greatly scrambled and almost flat, displaying convergence to the infinite temperature thermal state at fixed magnetization, i.e. $\propto e^{-\mu \hat{S}^z}$ for a suitably chosen chemical potential $\mu$. See Fig. \ref{s_fig_entropy} for the entropy growth.
}
\end{figure}

\be
m^\text{dr}_j(\pm\infty)=m_j-\sum_i t_{j,i} \sigma_i \vartheta_i(\pm\infty,\Phi)m^\text{dr}_i(\pm\infty)\,, \hspace{3pc}t_{j,i}=\int_{-\infty}^{+\infty}\dd\lambda\, T_{j,i}(\lambda)\, .
\ee
This allows to compute the dressing magnetizations without leaving the $\tau$ space
\be
m^\text{dr}_j(+\infty)=m_j-\sum_i t_{j,i} \sigma_i \tilde{\vartheta}_i(L_i,\Phi)m^\text{dr}_i(+\infty)\,,\hspace{2pc}m^\text{dr}_j(-\infty)=m_j-\sum_i t_{j,i} \sigma_i \tilde{\vartheta}_i(0,\Phi)m^\text{dr}_i(-\infty)\, .
\ee

Let us now discuss how to properly implement the boundary conditions. For each boundary in the $\tau$ space, we divide the fillings in incoming and outgoing accordingly to the shift dictated by the hydrodynamic equation. Assuming $\Phi$ is increased, at the left boundary we define a filling as incoming if $\sigma_i m_i^\text{dr}(-\infty)<0$, ourgoing otherwise. Instead, for the other boundary (which corresponds to $\tau=L_j$ for each string) we say a filling is incoming if $\sigma_i m_i^\text{dr}(+\infty)>0$, outgoing otherwise. If the flux is decreased, the definition of incoming and outgoing fillings are swapped.

Then, since at each boundary the value of the incoming fillings is fixed by the boundary content, the free degrees of freedom we can play with are only the outgoing fillings. Thus, at each boundary we keep fixed the incoming fillings, while we select the outgoing ones in such a way the related constraint \eqref{s_eq_magflux} is satisfied and the entropy creation rate \eqref{s_eq_entropy_rate} maximized.

\begin{figure}[t!]
\includegraphics[width=0.5\textwidth]{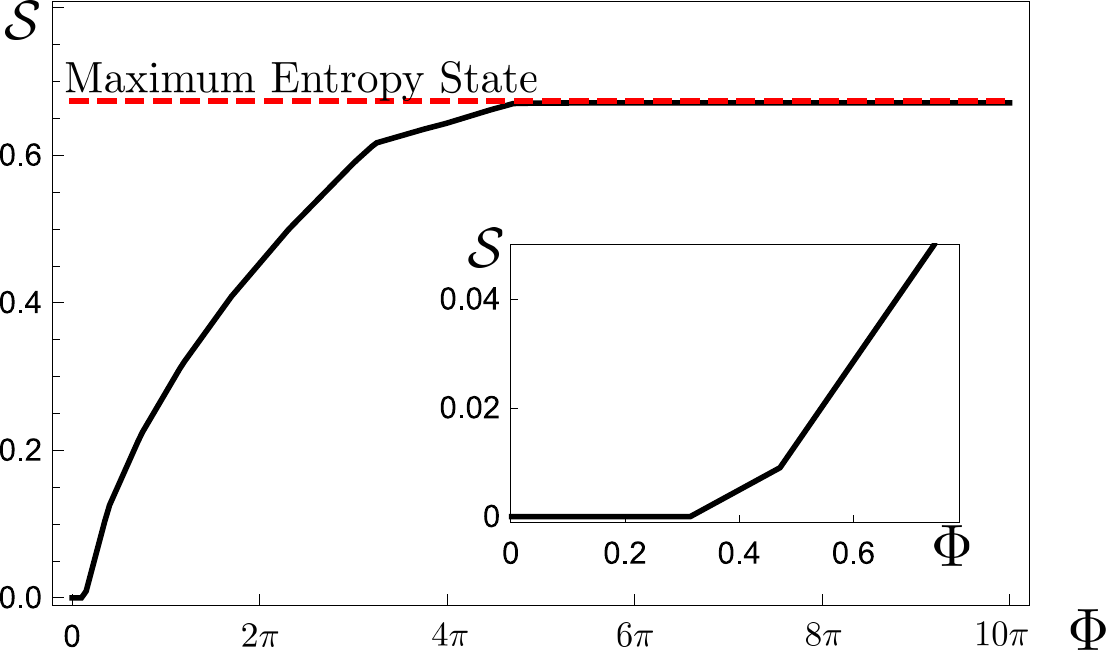} 
\caption{\label{s_fig_entropy}Entropy evolution for $\Delta=0.5$ and $\langle \hat{s}^z_j\rangle=0.1$ (see also Fig. \ref{fig_filling_deltaless}). The flux is steadily increased until $\Phi=10\pi$: for large fluxes the entropy saturates to that of the infinite temperature ensemble $\propto e^{-\mu \hat{S}^z}$, which has maximum entropy. For small values of the flux (see inset) the entropy does not grow until a critical value: this is the value of the flux where the $1^\text{st}$ string meets for the first time the boundary  (see Fig. \ref{fig_filling_deltaless}) and its value is state dependent. If the flux is changed within this critical value, the dynamics is reversible: the reversibility is broken as soon as the critical value is overcome. }
\end{figure}

The hydrodynamic equations are numerically solved in the $\tau$ space writing Eq. \eqref{s_eq_hydrotau} in the form of an infinitesimal shift, paying attention to suitably evolve $L_j$ as well. Once the hydrodynamic has been solved in the $\tau$ space, the change of coordinates Eq. \eqref{s_eq_tauvar} is inverted by mean of an iterative scheme, then the expectation value of the desired charges and currents can be computed. In Fig. \ref{fig_filling_deltaless} we depict the evolution of the root densities for the case $\Delta=1/2$ and constantly increasing the flux, being the system initialized in the ground state. In Fig. \eqref{s_fig_entropy} we show the entropy density growth for the same protocol.

\subsection{Periodicity of the $|\Delta| \geq 1$ single-string states}
\label{sec_periodic}

In this section, we consider $|\Delta| \geq 1$ and we show that, when the system is initially prepared in the groundstate at arbitrary filling, the dynamics is exactly periodic for $\Phi \to \Phi + 2\pi$, in agreement with what was observed both from the numerical solution of the hydrodynamic equations and from the simulation of the quantum spin-chain dynamics. More generally, we consider initial eigenstates for which only one filling function $\vartheta_k(\lambda)$ is non-vanishing
\begin{equation}
\label{singlestring}
\vartheta_j(\lambda) = \delta_{jk} \vartheta_k(\lambda)  \;, \qquad j = 1,2,\ldots
\end{equation}
As all the fillings except $j = k$ vanish, the hydrodynamic equation \eqref{s_eq_hydro_fill} is non-trivial only for $\vartheta_k(\lambda, \Phi)$.
Similarly to what was done in for $|\Delta| \leq 1$ in Eq. \eqref{s_eq_tauvar}, we can introduce the new parametrization in terms of
\be\label{s_eq_tauvarDeltageq1}
\tau_k(\lambda)=2\pi \int_{\lambda_0}^\lambda\dd\mu\, \rho_k^t(\mu,\Phi)\, .
\ee
where in this case the lower extreme $\lambda_0$ of the integral can be chosen arbitrarily as rapidities live on the compact range $[-\pi/2,\pi/2]$. Repeating the steps which led to \eqref{s_eq_hydrotau}, one obtains 
\be\label{s_eq_hydrotau_large}
 \tilde{\vartheta}_k(\tau,\Phi + \dd\Phi) \simeq  \tilde{\vartheta}_k(\tau - m^\text{dr}_k(\lambda_0, \Phi) \dd\Phi,\Phi)\, , 
\ee
i.e., in the variable $\tau$, the function $\tilde{\vartheta}_k(\tau, \Phi)$ shifts with an instantaneous velocity $m^\text{dr}_k(\lambda_0, \Phi)$. Since the initial state is periodic $\vartheta_k(\lambda + \pi) = \vartheta_k(\lambda)$, the initial state is re-obtained for a value $\Phi_0$ satisfying
\begin{equation}
\label{periodeq}
\int_0^{\Phi_0} \dd\Phi \, m^\text{dr}_k(\lambda_0, \Phi) =L_k 
\end{equation}
with
\be
L_k=2\pi \int_{-\pi/2}^{\pi/2} \dd\mu\, \rho_k^t(\mu,\Phi)\, .
\ee
From the hydrodynamic equations, we immediately get that $L_k$ is constant $\partial_\Phi L_k=0$.
That the left-hand side of Eq. \eqref{periodeq} does not depend on $\lambda_0$.
Furthermore, we now show that the value of the period $\Phi$ is independent on the specific details of the initial state $\vartheta_k(\lambda)$. Indeed, from \eqref{dresssuppl} and \eqref{singlestring}, we have
\begin{multline}
\label{mdresssingle}
m^\text{dr}_k(\lambda_0, \Phi)=k-\int_{-\pi/2}^{\pi/2} \dd\mu \, T_{k,k}(\lambda_0-\mu)\vartheta_{k}(\mu, \Phi)m_{k}^\text{dr}(\mu, \Phi) = \\=
k- \Omega_{k,k}(\pi/2) \vartheta_{k}(\lambda_0, \Phi)m_{k}^\text{dr}(\lambda_0, \Phi) - 
\int_{\lambda}^{\lambda+\pi} \dd\mu \,  \Omega_{k,k}(\lambda_0-\mu)\partial_\mu(\vartheta_{k}(\mu, \Phi)m_{k}^\text{dr}(\mu, \Phi) )
\, .
\end{multline}
where we set $\Omega_{j,k}(\lambda) = \int_{-\pi/2}^\lambda T_{j,k}(\lambda)$ and used integration by part. Using the hydrodynamic equation \eqref{hydrorho} and the periodicity $\rho_k(\lambda, \Phi_0) = \rho_k(\lambda, 0)$, we can integrate over $\Phi$ and obtain
\be\label{mdressint}
\int_0^{\Phi_0} \dd\Phi \, m^\text{dr}_k(\lambda_0, \Phi) = \Phi_0 k -  (2k-1)\int_0^{\Phi_0} \dd\Phi \, \vartheta_{k}(\lambda_0, \Phi)m_{k}^\text{dr}(\lambda_0, \Phi) 
\ee
where we used that $\Omega_{kk}(\pi) = 2k-1$.
Similarly, from \eqref{rhoteq}, we get
\be\label{eq_nfr}
2\pi \int_{-\pi/2}^{\pi/2}\dd\lambda \,\rho_k^t(\lambda, \Phi)=2\pi - 2\pi(2k-1)\int_{-\pi/2}^{\pi/2}\dd\mu\, \rho_k(\mu) = 2\pi ( 1 - (2k -1)\mathfrak{n}/k)\, ,
\ee
where we used that $\int_{-\pi/2}^{\pi/2} \dd\mu\, f_k(\mu) = 1$ and $\mathfrak{n}$ represents the filling of the state. Inverting \eqref{rhoteq}, we can write $\rho_k^t = (1 + T_{kk} \vartheta_k)^{-1} \ast f_k$.
This implies
\be
\mathfrak{n} = k\int_{-\pi/2}^{\pi/2} \dd\mu\, \rho_k(\mu, \Phi)  = k \int_{-\pi/2}^{\pi/2} \dd\mu \,   (\vartheta_k^{-1} + T_{kk} )^{-1} \ast f_k = \int_{-\pi/2}^{\pi/2} \dd\mu\, f_k (\mu) \vartheta_k(\mu, \Phi) 
m^\text{dr}_k(\mu, \Phi)\, ,
\ee
where in the last equality we used that  $[(\vartheta_k^{-1} + T_{kk} )^{-1} \ast k] (\mu) = \vartheta_k(\mu, \Phi) m^\text{dr}_k(\mu, \Phi)$. 
Using the $\lambda_0-$independence of Eq. \eqref{mdressint} (and again $\int_{-\pi/2}^{\pi/2} \dd\mu\, f_k(\mu) = 1$), we can equivalently write
\be
\int_0^{\Phi_0} \dd\Phi \, m^\text{dr}_k(\lambda_0, \Phi) = \Phi_0 k -  (2k-1)\int_0^{\Phi_0} \dd\Phi\int_{-\pi/2}^{\pi/2}\dd\lambda \, f_k(\lambda) \vartheta_{k}(\lambda, \Phi)m_{k}^\text{dr}(\lambda_0, \Phi) = \Phi_0 k -  (2k-1)\int_0^{\Phi_0} \dd\Phi\,\mathfrak{n} \, .
\ee
Since $L_k$ is $\Phi-$indepentent, $\mathfrak{n}$ is flux independent as well as evident from Eq. \eqref{eq_nfr}. Therefore, the flux integration can be simply carried out
\be
\int_0^{\Phi_0} \dd\Phi \, m^\text{dr}_k(\lambda_0, \Phi) =\Phi_0\left( k -  (2k-1)\mathfrak{n} \right)\, .
\ee
Combining the above with Eq. \eqref{periodeq} with Eq. \eqref{eq_nfr} we finally get
\be
\Phi_0  = \frac{2\pi}{k} \;.
\ee

\section{Numerical methods}
\label{sec_num}
In this section, we provide details of the numerics presented in the main text. We employed two different methods: exact diagonalization (ED) for finite systems and time-dependent variational principle (TDVP)~\cite{TDVP} for the system in the thermodynamic limit. 
\begin{itemize}
\item ED allows reaching arbitrary times but has strong limitation on the system size because of the exponential growth of the Hilbert space dimension in the number of spins $N$. We reduced the computational effort using the two symmetries of the problem, $U(1)$ charge symmetry and invariance under translation, to project into the sector of fixed magnetization and momentum which contains the groundstate. In this way, we were able to access as maximal sizes $N = 25$ for $\langle s_z^j \rangle = 0.1$ and $N = 50$ for $\langle s_z^j \rangle = 0.4$. In order to simulate the quantum dynamics, we employed the fourth-order Runge-Kutta method to solve the many-body Schr\"odinger equation with a time-dependent flux. 
\item TDVP is based on representing the quantum state as a \textit{matrix-product state} with fixed bond dimension $\chi$ and considering the Schr\"odinger equation projected in the MPS manifold~\cite{TDVP}. In our approach, the resulting equation was again solved with Runge-Kutta method. We considered $\chi = 128$ and $\chi = 256$ obtaining similar results for all regimes were entanglement entropy production is suppressed. When entropy is generated, the bond dimension is saturated exponentially fast in time and error is accumulated: it is therefore harder to access slow flux variations. In the main text, we only showed data where data for $\chi = 128$ and $\chi = 256$ do not show significant differences. 
\end{itemize}

\end{document}